\documentclass[iop]{emulateapj}

\usepackage{amsmath}

\def\vec#1{\mbox{\boldmath $#1$}}

\slugcomment{Draft version March 23, 2017}

\shorttitle{Distribution of captured planetesimals in circumplanetary gas disks}
\shortauthors{Suetsugu \& Ohtsuki}

\begin{document}


\title{Distribution of captured planetesimals in circumplanetary gas disks and implications for accretion of regular satellites}


\author{Ryo Suetsugu\altaffilmark{1}, Keiji Ohtsuki\altaffilmark{2}}
\affil{1.School of Medicine, University of Occupational and Environmental Health, Kita-Kyushu,
807-8555, Japan}
\affil{2.Department of Planetology, Kobe University, Kobe 657-8501, Japan}

\email{ryos@med.uoeh-u.ac.jp}



\begin{abstract}
Regular satellites of giant planets are formed by accretion of solid bodies in circumplanetary disks. 
Planetesimals that are moving on heliocentric orbits and are sufficiently large to be decoupled 
from the flow of the protoplanetary gas disk can be captured by gas drag from the circumplanetary disk. 
In the present work, we examine the distribution of captured planetesimals in circumplanetary disks using orbital integrations. 
We find that the number of captured planetesimals reaches an equilibrium state as a balance between continuous capture and orbital decay into the planet. 
The number of planetesimals captured into retrograde orbits is much smaller than those on prograde orbits, 
because the former ones experience strong headwind and spiral into the planet rapidly. 
We find that the surface number density of planetesimals at the current radial location of regular satellites can be significantly enhanced by gas drag capture, depending on the velocity dispersions of planetesimals and the width of the gap in the protoplanetary disk. 
Using a simple model, we also examine the ratio of the surface densities of dust and captured planetesimals in the circumplanetary disk, 
and find that solid material at the current location of regular satellites can be dominated by captured planetesimals when the velocity dispersion of planetesimals is rather small and a wide gap is not formed in the protoplanetary disk. 
In this case, captured planetesimals in such a region can grow by mutual collision before spiraling into the planet, and would contribute to the growth of regular satellites.
\end{abstract}




\section{INTRODUCTION}
\label{sec:intro}

Regular satellites of giant planets have nearly circular and coplanar prograde orbits, and are thought to have formed in  circumplanetary disks \citep{CW09, E09}.
Gas and solids are supplied into the circumplanetary disk from the vicinity of the giant planet's orbit in the protoplanetary disk. 
\citet{CW02} proposed the so-called gas-starved disk model for the formation of regular satellites of giant planets, where the satellites are formed in a waning circumplanetary disk at the very end stage of giant planet formation.
\citet{CW06} performed N-body simulation of satellite formation based on the above model, and showed that mass fraction of satellite system relative to the host planet is regulated to $\sim10^{-4}$, which is consistent with observations. Later studies revised this model and tried to explain the difference between the satellite systems of Jupiter and Saturn  \citep{S10, OI12}. 
On the other hand, the so-called solid enhanced minimum mass disk model proposed by \citet{ME03} considers a circumplanetary disk consisting of two different parts of surface density in order to explain the rock-ice ratio of  the Galilean satellites.
Recently, satellite formation based on this model was examined semi-analytically \citep{MI16}.
Although growth and orbital evolution of satellites  in the circumplanetary disk have been investigated in detail by these works,
the distribution of solid materials that are building blocks of satellites in circumplanetary disks has been poorly understood.

There are two kinds of processes for the supply of solid materials into circumplanetary disks. 
If incoming bodies are sufficiently small to be coupled with the gas flow, 
they are brought to the circumplanetary disk with the gas inflow from the protoplanetary disk \citep{CW02}.
\citet{T14} examined capture process of solid materials with various sizes, using results of hydrodynamic simulations of gas flow around a growing giant planet \citep{M08, T12} and three-body orbital integration. 
They found that accretion efficiency of dust-size particles initially on circular orbits is rather low when the gas surface density of the protoplanetary disk is that of the  minimum mass nebula model.
They also found that such small solid bodies can be supplied to the circumplanetary disk when the gas surface density becomes smaller as a results of disk dispersal.
Another process for the supply of solid bodies is  capture of planetesimals by gas drag from the circumplanetary disk \citep{F13, T14, DP15}.
In the case of large planetesimals that are decoupled from the gas flow, effects of gas drag on their orbits become significant only when they pass through the dense part of the circumplanetary disk in the vicinity of the planet.
\citet{F13} performed three-body orbital integrations and examined capture of planetesimals from their heliocentric orbits into the circumplanetary disk. 
They found that planetesimals approaching the circumplanetary disk in the retrograde direction (i.e., in the direction opposite to the motion of the gas in the disk) are more easily captured by gas drag owing to the larger velocity relative to the gas. 
They also found that capture rates decrease with increasing size of  planetesimals because strong gas drag is needed to capture large ones \citep[see also][]{T14}.
\citet{DP15} examined orbital evolution of planetesimals in the vicinity of Jupiter's orbit, using global orbital integration including gas drag, disruption via ram pressure, and mass loss through ablation.
They found that planetesimals are captured from both interior and exterior orbits.
Although they also show the distribution of orbital elements of captured planetesimals in the circumjovian disk, they did not examine their orbital evolution in the disk.

Recently, we examined the capture and subsequent orbital evolution of planetesimals that are decoupled from the gas flow in circumplanetary gas disks using three-body orbital integration.
In \citet{S16}, we investigated the orbital evolution and the distribution of orbital elements of planetesimals captured by relatively strong gas drag, where planetesimals lose sufficient amount of energy to become captured in a single encounter with the planet.
On the other hand, in \citet{SO16}, we examined the capture and subsequent orbital evolution of irregular satellites in waning circumplanetary gas disks.
We found that irregular satellites are easily captured and are likely to survive when  the gas drag is strong and the disk dispersal takes place in a short timescale.
However, in these works, radial distribution of solid materials in the circumplanetary disk was not studied in detail. 
Furthermore, the abundance of captured planetesimal-size bodies relative to dust-size particles  in the circumplanetary disk is important to understand which solid materials mainly contribute to the growth of regular satellites.
The distribution and size of solid materials in the circumplanetary disk would  also influence the timescale of satellite accretion and their chemical composition such as the ice-rock ratio \citep{SG12, D13}.

In the present work, as an extension of our recent works \citep{S16, SO16}, we examine radial distribution of planetesimals captured in circumplanetary disks.
We perform orbital integration for planetesimals that are decoupled from the gas inflow but are affected by gas drag from the circumplanetary gas disk. 
We examine capture of planetesimals from their heliocentric orbits by gas drag from the circumplanetary gas disk and orbital evolution of captured planetesimals in the disk. 
Using our numerical results, we obtain distribution of the surface number density of captured planetesimals in the circumplanetary disk. 
In Section~\ref{sec:num}, we describe basic equations, disk model, and numerical methods used in the present work. 
We show dynamical evolution and distribution of planetesimals captured from heliocentric circular orbits in Section~\ref{sec:cir}. 
In Section~\ref{sec:eigap}, we investigate the effect of the dynamical states of planetesimals on the surface number density.
In Section~\ref{sec:dis}, we discuss implications of our numerical results for accretion of regular satellites.
Section~\ref{sec:sum} summarizes our results.

\section{THE MODEL}
\label{sec:num}
\subsection{Numerical Procedures}
\begin{figure}
\plotone{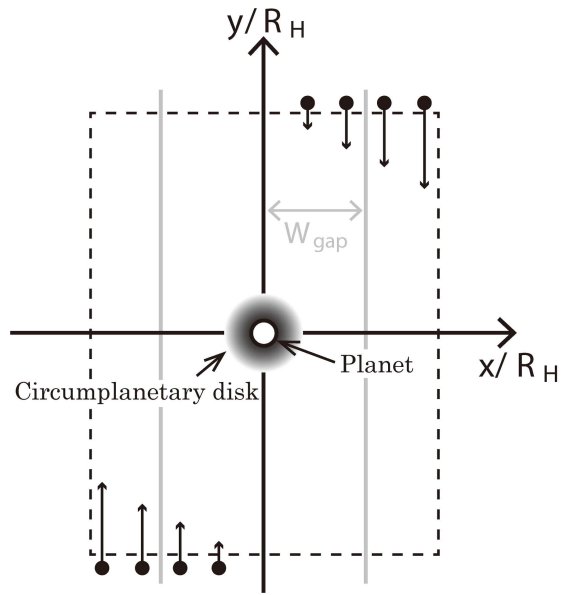}  
 \caption{Schematic illustration of our numerical setting.
A planet is located at the origin of the rectangular simulation cell, and has an axisymmetric thin circumplanetary disk. 
Planetesimals initially on unperturbed orbits are supplied through the azimuthal boundaries to the simulation cell, 
and their orbits are numerically integrated taking account of the gravity of the planet and gas drag from the circumplanetary disk. 
Mutual gravity between planetesimals is neglected. 
In the case of non-uniform radial distribution of planetesimals (Section~\ref{sec:eigap}), planetesimals in the vicinity of the planet's orbit are removed, 
forming a gap centered on the planet's orbit with a half width measured in units of $R_{\rm H}$ being $W_{\rm gap}$.}
 \label{fig:simu}
\end{figure}
We consider a local coordinate system ($x, y, z$) centered on a planet (Figure~\ref{fig:simu}).
We assume that the planet ($M$) is on a circular orbit with semi-major axis $a_{0}$, and is embedded in a disk of planetesimals. 
Also, the planet is assumed to have  a circumplanetary gas disk, whose mid-plane coincides with the planet's orbital plane. 
In the present work, we assume a disk of equal-mass planetesimals ($m_{\rm s}\ll M$), and neglect gravitational interaction among them. 
At azimuthal locations far from the planet, planetesimals are assumed to have uniform radial distribution.
Planetesimals are supplied through the azimuthal boundaries at $y = \pm 50 R_{\rm H}$ ($R_{\rm H}$ is the planet's Hill radius; $R_{\rm H}=a_{0}h_{\rm H}=a_{0}\{\left(M+m_{\rm s}\right)/(3M_{\odot})\}^{1/3}$), which is sufficiently far from the planet to neglect its gravitational effect. 
Supplied planetesimals are assumed to be distributed radially uniformly in the protoplanetary disk; we examine the effect of non-uniform radial distribution in Section \ref{sec:eigap}.
We integrate the orbits of planetesimals by solving Hill's equation with the effect of gas drag from the circumplanetary disk.
The motion of the planetesimal is expressed as \citep{O12, SO16},
\begin{eqnarray}
\ddot{x}&=&2\Omega\dot{y}+3\Omega^{2}x-\frac{G\left(M+m_{\rm s}\right)}{R^{3}}x+a_{{\rm drag}, x}, \nonumber \\
\ddot{y}&=&-2\Omega\dot{x}-\frac{G\left(M+m_{\rm s}\right)}{R^{3}}y+a_{{\rm drag}, y}, \nonumber \\
\ddot{z}&=&-\Omega^{2}z-\frac{G\left(M+m_{\rm s}\right)}{R^{3}}z+a_{{\rm drag}, z},
\label{eq:hillgas}
\end{eqnarray}
where $\Omega$ is the planet's orbital angular frequency, and $R=\sqrt{x^{2}+y^{2}+z^{2}}$ is the distance between the centers of the planet and the planetesimal.
$\vec{a}_{\rm drag}\equiv \vec{F}_{\rm drag}/m_{\rm s}$ is the acceleration due to the gas drag force $\vec{F}_{\rm drag}$ given by
\begin{eqnarray}
\vec{F}_{\rm drag}=-\frac{1}{2}C_{\rm D}\pi r^{2}_{\rm s}\rho_{\rm gas}u\vec{u},
\end{eqnarray}
where $C_{\rm D}$ is the drag coefficient (we assume $C_{\rm D}=1$; Appendix A), $r_{\rm s}$ is the radius of the planetesimal, 
$\rho_{\rm gas}$ is the gas density, and $u=|\vec{u}|$ is the velocity of the planetesimal relative to the gas.
We scale the distance by mutual Hill radius $R_{\rm H}$ and time by $\Omega^{-1}$.
Then, we can express the above equation of motion in a non-dimensional form as
\begin{eqnarray}
\ddot{\tilde{x}}&=&2\dot{\tilde{y}}+3\tilde{x}-\frac{3\tilde{x}}{\tilde{R}^{3}}+\tilde{a}_{{\rm drag}, x}, \nonumber \\
\ddot{\tilde{y}}&=&-2\dot{\tilde{x}}-\frac{3\tilde{y}}{\tilde{R}^{3}}+\tilde{a}_{{\rm drag}, y},  \\
\ddot{\tilde{z}}&=&-\tilde{z}-\frac{3\tilde{z}}{\tilde{R}^{3}}+\tilde{a}_{{\rm drag}, z}, \nonumber
\label{eq:hillgas_non}
\end{eqnarray}
where tildes denote non-dimensional quantities.
The non-dimensional acceleration due to gas drag can be written as
\begin{eqnarray}
\vec{\tilde{a}}_{\rm drag}=\vec{a}_{\rm drag}/(R_{\rm H}\Omega^{2})=-\frac{3}{8}C_{\rm D}\frac{\rho_{\rm gas}}{\tilde{r}_{\rm s}\rho_{\rm s}}\tilde{u}\vec{\tilde{u}},
\label{eq:nga}
\end{eqnarray}
where $\rho_{\rm s}$ is the internal density of planetesimals.
When the gas drag can be neglected, we can confirm from Equation~(\ref{eq:hillgas_non}) the conservation of energy given as
\begin{eqnarray}
E=\frac{1}{2}\left(\dot{\tilde{x}}^{2}+\dot{\tilde{y}}^{2}+\dot{\tilde{z}}^{2}\right)+U(\tilde{x}, \tilde{y}, \tilde{z}),
\label{eq:enegas} 
\end{eqnarray}
with
\begin{eqnarray}
U(\tilde{x}, \tilde{y}, \tilde{z})=-\frac{1}{2}\left(3\tilde{x}^{2}-\tilde{z}^{2}\right)-\frac{3}{\tilde{R}}+\frac{9}{2}.
\label{eq:potgas}
\end{eqnarray}
We remove planetesimals when their distance from the planet becomes sufficiently large or when collision with the planet is detected. 
We assume that the physical size of the planet ($r_{\rm p}$) relative to its Hill radius, $r_{\rm p}/R_{\rm H}$, is $10^{-3}$, 
which corresponds to the physical size of a planet at Jupiter's orbit.

The number of planetesimals in the circumplanetary disk increases due to capture by gas drag and decreases due to collision with the planet after orbital decay.
However, as we will show below,  the number reaches an equilibrium state by a balance between continuous capture of planetesimals and loss to the planet due to orbital decay.
Since the timescale needed to reach a quasi-steady state depends on the dynamical properties of planetesimals and the strength of gas drag,   
we perform orbital integration for a sufficiently long period of time ($200-500T_{\rm K}$, where $T_{\rm K}$ is the orbital period of the planet).

Unperturbed surface number density of the planetesimal disk $n_{\rm s}$ is one of the important parameters in order to determine the distribution of captured planetesimals in the circumplanetary disk.
Thus, we briefly estimate the value of $n_{\rm s}$.
We can obtain $n_{\rm s}$ when we set the solid surface density in the protoplanetary disk $\Sigma_{\rm ppd, solid}$ and the mass of the planetesimal $m_{\rm s}$ (recall that we assume equal-sized planetesimals).
If we adopt the minimum mass solar nebula model \citep{H81}, the surface density of solids in the protoplanetary disk is $\Sigma_{\rm ppd, solid}\simeq3.3$ gcm$^{-2}\simeq9\times10^{25}$ g$R_{\rm H, J}^{-2}$ at the orbit of Jupiter ($R_{\rm H, J}$ is the Hill radius of Jupiter).
In the present work, we will examine the distribution of captured planetesimals in circumplanetary disks for cases with a wide range of planetesimal sizes ($r_{\rm s}\sim1-10^{4}$ m).
If we assume that the size of the planetesimal  is $r_{\rm s}=300$ m,    
the surface number density scaled by the Hill radius of Jupiter can be written by $\tilde{n}_{\rm s}=8.2\times10^{11}\left(\rho_{\rm s}/1 {\rm gcm^{-3}}\right)^{-1}\left(r_{\rm s}/300 {\rm m}\right)^{-3}$ $R_{\rm H, J}^{-2}$.
However, if we adopt such a realistic surface number density, the number of  orbits to integrate per one timestep is $\sim10^{14}$ 
(Even if we assume a large size of planetesimals to reduce the surface number density to be $\tilde{n}_{\rm s}=8.2\times10^{5}$, we have to integrate $\sim10^{8}$ orbits).
Here, we should recall that we neglect mutual gravity between planetesimals, thus their orbits are not influenced by the choice of $\tilde{n}_{\rm s}$. 
Therefore, we adopt a value of $\tilde{n}_{\rm s}$ feasible for our simulation (typically, $\tilde{n}_{\rm s}=400$), and discuss the enhancement of the surface number density in the circumplanetary disk relative to the background value
\footnote{In this case, the number of planetesimals in the simulation cell is $\sim10^{4}$ for the case of planetesimals initially on circular orbits, and  is $\sim10^{5}$ orbits for the case of planetesimals initially on eccentric and inclined orbits.
We confirmed that the surface number density of captured planetesimals in the circumplanetary disk is unchanged when $\tilde{n}_{\rm s}$ is increased to 1000.
Therefore, the above value of $\tilde{n}_{\rm s}$ is sufficiently large to accurately obtain the distribution of captured planetesimals in the circumplanetary disk.}.

\subsection{Gas Drag}

For those planetesimals that are large enough to be decoupled from the inflowing gas,
gas drag from the circumplanetary disk becomes important only when they pass through the dense part of the disk in the vicinity of the planet, 
where the gas distribution can be approximated to be axisymmetric \citep{T12, F13}. 
Thus, in the present work, we assume that the radial distribution of the gas density of the circumplanetary disk is axisymmetric, its radial dependence is given by a power law, and its vertical structure is isothermal.
Under these assumptions, the gas density can be written by 
\begin{eqnarray}
\rho_{\rm gas}=\frac{\Sigma}{\sqrt{2\pi}h}{\rm exp}\left( -\frac{z^{2}}{2h^{2}}\right),
\label{eq:rhodis}
\end{eqnarray}
where $h=c_{\rm s}/\Omega_{\rm p}$ is the scale height of the circumplanetary disk ($\Omega_{\rm p}$ is the Keplerian orbital frequency around the planet), and
\begin{eqnarray}
\Sigma = \Sigma_{\rm rd}\left(\frac{r}{r_{\rm d}}\right)^{-p}, \;\;\;\;\; c_{\rm s}=c_{\rm rd}\left(\frac{r}{r_{\rm d}}\right)^{-q/2}
\label{eq:snum_svelo}
\end{eqnarray}
are the gas surface density and sound velocity, respectively, with $r=\sqrt{x^{2}+y^{2}}$ being the horizontal distance from the planet in the mid-plane.
In the above, $r_{\rm d}=dR_{\rm H}$ is a typical length scale roughly corresponding to the effective size of the circumplanetary disk, 
and $\Sigma_{\rm rd}$ and $c_{\rm rd}$ are the surface density and sound velocity at that radial location. 
In our calculations, we set $d=0.2$ and $p=3/2$ based on results of hydrodynamic simulations \citep{M08, T12}, and also assume $q=1/2$ as a simple model \citep{F13, SO16, S16}. 
In our orbital calculations we turn on gas drag when planetesimals enter the planet's Hill sphere, in order to avoid effects of artificial cutoff at $r=r_{\rm d}$. 
Because the gas density decreases rapidly with increasing distance from the planet, this assumption on gas drag does not affect results of our calculations\footnote{For those small planetesimals whose motion is strongly affected by the gas flow, the effect of gas drag is important even outside of the planet's Hill sphere \citep{T14}. 
However, calculations of long-term evolution of captured bodies including such an effect are time-consuming, 
and in the present work, we focus on the orbital evolution under gas drag within the Hill sphere, by making the above assumption.}.
Gas elements in the disk are assumed to rotate in circular orbits around the planet with a velocity slightly lower than the Keplerian velocity due to radial pressure gradient, i.e.,
\begin{eqnarray}
v_{\rm gas}=(1-\eta)v_{\rm K},
\end{eqnarray}
where $v_{\rm K}$ is the Keplerian velocity around the planet at the radial location considered.
Using Equations (\ref{eq:rhodis}) and (\ref{eq:snum_svelo}), $\eta$ can be written as \citep{T02}
\begin{eqnarray}
\eta = \frac{1}{2}\frac{h^{2}}{r^{2}}\left(p+\frac{q+3}{2}+\frac{q}{2}\frac{z^{2}}{h^{2}}\right).
\end{eqnarray}

When the gas density is given by Equation (\ref{eq:rhodis}), Equation (\ref{eq:nga}) can be rewritten as \citep{F13}
\begin{eqnarray}
\vec{\tilde{a}}_{\rm drag}=-\zeta\tilde{r}^{-\gamma}{\rm exp}\left(-\frac{\tilde{z}^{2}}{2\tilde{h}^{2}}\right)\tilde{u}\vec{\tilde{u}},
\label{eq:dragterm}
\end{eqnarray}
where $h=h_{\rm rd}\left(r/r_{\rm d}\right)^{(3-q)/2}$ with $h_{\rm rd}$ being the scale height at $r=r_{\rm d}$, and $\gamma\equiv p+(3-q)/2$, and $\zeta$ is the non-dimensional parameter representing the strength of gas drag defined by
\begin{eqnarray}
\begin{split}
\zeta&\equiv \frac{3}{8\sqrt{2\pi}}\frac{C_{\rm D}}{r_{\rm s}\rho_{\rm s}}\frac{\Sigma_{\rm rd}}{\tilde{h}_{\rm rd}}d^{\gamma} \\
      &=3 \times 10^{-7} C_{\rm D} \left( \frac{r_{\rm s}}{1 {\rm km}} \right)^{-1} \left( \frac{\rho_{\rm s}}{1 {\rm g\, cm^{-3}}} \right)^{-1} \\ 
      &\quad \times \left( \frac{\Sigma_{\rm rd}}{1 {\rm g\, cm^{-2}}} \right) \left( \frac{\tilde{h}_{\rm rd}}{0.06} \right)^{-1} \left( \frac{d}{0.2} \right)^\gamma. 
          \end{split}
            \label{eq:gaszeta}
\end{eqnarray}
We set $\tilde{h}_{\rm rd}\equiv h_{\rm rd}/R_{\rm H}=0.06$ in the present work \citep{T12, F13}
\footnote{In the following, we assume icy planetesimals with $\rho_{\rm s} = 1$gcm$^{-3}$ when converting $\zeta$ into $r_{\rm s}$. 
The value of $\zeta$ for rocky planetesimals becomes smaller than that for icy planetesimals with the same size. 
This leads to smaller capture rates, while the lifetime of captured bodies in the circumplanetary disk becomes longer. 
As a result, the influence of the material density on the radial distribution of captured bodies is expected to be small.}.
If we assume the gas surface density based on the gas-starved disk model \citep{CW02},  we have $\Sigma_{\rm rd}\simeq1$gcm$^{-2}$.
For the above fiducial values of  $C_{\rm D}$, $d$,  and $\tilde{h}_{\rm rd}$, 
the relation between $\zeta$ and the planetesimal size $r_{\rm s}$ can be written as \citep[see][]{F13, SO16}
\begin{eqnarray}
r_{\rm s}\simeq 0.03\left(\frac{\Sigma_{\rm rd}}{1 {\rm g\, cm}^{-2}}\right)\left(\frac{\rho_{\rm s}}{1 {\rm g\, cm}^{-3}}\right) ^{-1}\zeta^{-1}\;\;\;\;\; {\rm cm}.
\label{eq:size}
\end{eqnarray}
In the present work, we will examine cases with $\zeta=10^{-8}-10^{-4}$, which roughly corresponds to planetesimal
sizes on the order of $r_{\rm s}\sim1-10^{4}$ m.

\section{CASE OF INITIALLY HELIOCENTRIC CIRCULAR ORBITS}
\label{sec:cir}
\subsection{Supply of Planetesimals into Circumplanetary Disk}

\begin{figure*}
 \epsscale{1.15}
  \plotone{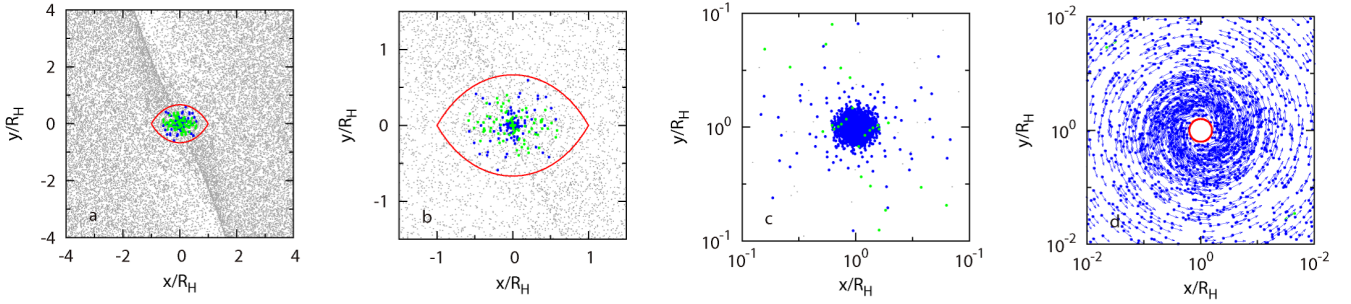}
  \epsscale{1}
\caption{Example of snapshots of the spatial distribution of planetesimals in different scales in the case of $\zeta=10^{-6}$ ($r_{\rm s}\simeq300$ m). 
Blue and green circles represent planetesimals captured in prograde orbits and retrograde orbits, respectively, while gray dots show free planetesimals. 
The lemon-shaped curves in Panels (a) and (b) show the planet's Hill sphere.
In Panel (d), vectors represent orbital velocity of the planet-centered orbits of captured planetesimals, and
the red circle shows the physical size of the planet.
}
 \label{fig:xy_flow}
\end{figure*}

\begin{figure}[ht]
\epsscale{1.2}
  \plotone{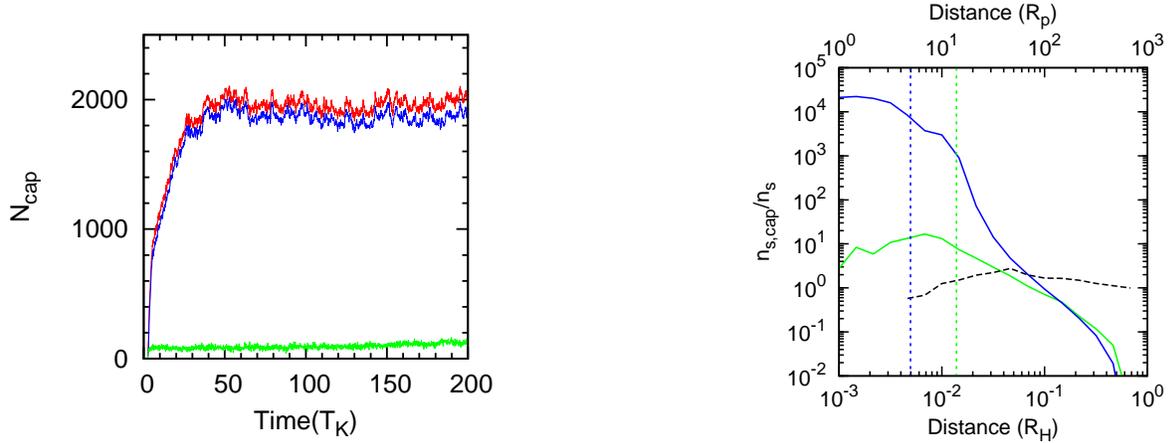}
  \epsscale{1}
 \caption{
Time variation of the number of planetesimals captured in the circumplanetary disk from their heliocentric circular orbits in the case of $\zeta=10^{-6}$ ($r_{\rm s}=300$ m).
 Red, blue, and green lines show the total number of captured planetesimals, and the numbers of planetesimals captured into prograde
and retrograde orbits, respectively.
}
 \label{fig:tncap_6}
\end{figure}

Figure~\ref{fig:xy_flow} shows an example of snapshots of the spatial distribution of planetesimals supplied from heliocentric circular orbits in the case of  $\zeta=10^{-6}$ ($r_{\rm s}\simeq300$ m).
Figure~\ref{fig:xy_flow}(a) shows the distribution of planetesimals in the vicinity of the planet.
Most of planetesimals are scattered by the gravity of the planet or pass by the planet's Hill sphere, while some of them enter the Hill sphere.
Planetesimals that enter the planet's Hill sphere are affected by gas drag from the circumplanetary gas disk, and those planetesimals that lose a sufficient amount of energy become permanently captured (Figure~\ref{fig:xy_flow}(b)).
In  Figure~\ref{fig:xy_flow}, blue and green marks show planetesimals permanently captured by gas drag ($E<0$);
blue and green represent those captured into prograde  and  retrograde orbits, respectively.
On the other hand, gray dots represent free planetesimals just passing by the planet.
Figure~\ref{fig:xy_flow}(c) and (d) show blow-ups of the distribution in the vicinity of the planet.
Typically, those captured into retrograde orbits tend to have larger velocity relative to the gas than those on prograde orbits. 
Thus, planetesimals in retrograde orbits spiral into the planet more quickly, 
and the innermost part of the circumplanetary disk is dominated by planetesimals captured into prograde orbits (Figure~\ref{fig:xy_flow}(d)).
Also, free planetesimals rarely exist in the vicinity of the planet, because they become permanently captured by strong gas drag in such a dense part of the disk.

\begin{figure}[ht]
\epsscale{1.2}
  \plotone{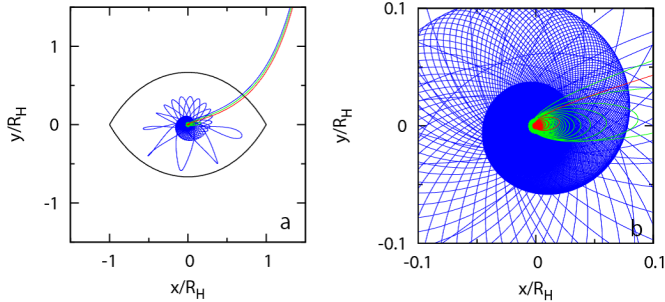}
\epsscale{1}
 \caption{(a) Examples of prograde capture orbits that have slightly different initial values of $b_{\rm H}=(a-a_{0})/R_{\rm H}$,
where $a$ and $a_{0}$ the semi-major axes of the planetesimal and the planet, respectively ($\zeta=10^{-6}$).
Blue, green, and red lines represent cases for $b_{\rm H}=2.054$, 2.062, and 2.070, respectively. 
The lemon-shaped curve in Panel (a) shows the planet's Hill sphere.
Panel (b) is the blow-up of Panel (a). 
 }
 \label{fig:cap_orbit}
\end{figure}

Figure~\ref{fig:tncap_6} shows evolution of the number of planetesimals permanently captured within the planet's Hill sphere for the case shown Figure~\ref{fig:xy_flow}.
The red curve represents the total number of captured planetesimals.
The number of captured planetesimals $N_{\rm cap}$ can be determined by a balance between the increase by the capture of planetesimals and the decrease due to collision with the planet after orbital decay.
We find that the total number first increases gradually  until  $t \simeq 35T_{\rm K}$ due to capture of planetesimals.
After $t\simeq35T_{\rm K}$, the system reaches an equilibrium state with $N_{\rm cap}\simeq2000$. 
In an unperturbed state without the planet and the circumplanetary disk,
the number of planetesimals passing within the planet's Hill sphere is $\tilde{n}_{\rm s}R_{\rm H}^{2}\pi\simeq1200$ when $\tilde{n}_{\rm s}=400$.
Also, the number of free planetesimals within the Hill sphere is about 1200 in this case. 
Thus, the number density of planetesimals within the Hill sphere is significantly enhanced due to the effects of the planet and the disk.
The blue and green lines represent the number of  planetesimals permanently captured in the prograde direction and the retrograde direction, respectively.
We find that the number of planetesimals captured in retrograde orbits is significantly smaller because of their rapid orbital decay into the planet.

\begin{figure}
\epsscale{1.2}
  \plotone{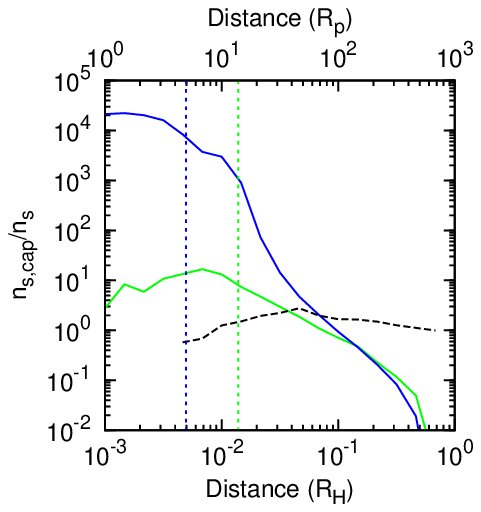}
  \epsscale{1}
 \caption{Surface number density of captured planetesimals in the case of $\zeta=10^{-6}$ ($r_{\rm s}\simeq300$ m) as a function of distance from the planet.  
The lower horizontal axis shows the distance in units of the Hill radius of the planet $R_{\rm H}$, while the upper axis is in units of the planetary radius $R_{\rm p}$.
Surface number density is scaled by the unperturbed value of the planetesimal disk $n_{\rm s}$. 
Blue and green lines show those  captured into prograde and retrograde orbits, respectively, while the black line shows that of free planetesimals.
Blue and green vertical dotted lines represent the prograde and retrograde capture radii, respectively \citep{F13}.
}
 \label{fig:rns_6}
\end{figure}

On close inspection of Figure~\ref{fig:tncap_6}, we find that the number of bodies captured in prograde orbits increases very rapidly at the very beginning ($t \lesssim 5 T_{\rm K}$), 
and then its growth rate slows down and continues gradual increase at $5T_{\rm K} \lesssim t \lesssim 35T_{\rm K}$. 
The rapid increase in the first stage is due to capture of bodies in a single encounter with the planet (Figure~\ref{fig:cap_orbit}; red and green lines). 
On the other hand, the gradual increase in the second stage comes from a different type of capture orbits, 
which maintain larger pericenter distance and survive longer in the disk (Figure~\ref{fig:cap_orbit}, blue line). 
Dynamical behavior of these orbits is described in more detail in Appendix B.


\subsection{Radial Distribution of Captured Planetesimals}

Figure~\ref{fig:rns_6} shows the surface number density of captured planetesimals $n_{\rm s, cap}$ scaled by the background surface number density $n_{\rm s}$, as a function of their radial distance from the planet.
These plots are created by averaging the distributions at different times after the system reached the equilibrium state.
The distribution of planetesimals captured into prograde orbits (blue line) show different slopes depending on the radial distance. 
In the outer region with $r/R_{\rm H} \gtrsim 0.04$, the distribution can be roughly approximated by a power law $n_{\rm s, cap} \propto r^{-\alpha}$ with $\alpha \simeq 2$. 
Then, with decreasing distance, the slope becomes steeper ($\alpha \simeq 4.5$) at $0.01 \lesssim r/R_{\rm H} \lesssim 0.04$, and then the distribution levels off at $r/R_{\rm H} \simeq 0.01$. 
With further decreasing distance, it increases again and, finally, levels off again at $r/R_{\rm H} \lesssim 0.003$. 
We find that the above behavior of the surface number density distribution reflects  dynamical evolution of captured planetesimals in the circumplanetary disk.
Those captured by a single encounter with the dense part of the gas disk accumulate in the vicinity of the planet and are short-lived because of rapid orbital decay, 
while those captured in the outer part are typically long-lived, with their pericenter avoiding penetration into the dense part of the disk. 
Captured bodies in the outer part ($r/R_{\rm H} \gtrsim 0.01$) are dominated by the latter long-lived ones. 
Immediately after being captured, their orbits decay rather rapidly due to large orbital eccentricities; this corresponds to the region with $\alpha \simeq 2$ ($r/R_{\rm H} \gtrsim 0.04$). Then, their eccentricities are damped by gas drag and orbits become nearly circular. 
Then the rate of orbital decay decreases \citep{A76, K15}; the region with the steeper slope with $\alpha \simeq 4.5$ ($0.01 \lesssim r/R_{\rm H} \lesssim 0.04$) corresponds to such slowly migrating bodies on the relatively long-lived orbits. 
With further decreasing radial distance, the rate of orbital decay increases because of higher gas density, which leads to the nearly flat distribution at $r/R_{\rm H} \lesssim 0.01$.

On the other hand, the increase of the surface number density in the inner region ($0.003\lesssim r/R_{\rm H}\lesssim0.007$) is caused by planetesimals captured into short-lived orbits (the red and green lines in Figure~\ref{fig:cap_orbit}).  
In fact, the outer boundary of this region roughly corresponds to the prograde capture radius where planetesimals can be captured by a single passage through the gas disk \citep{F13}. 
Since such captured planetesimals spiral into the planet quickly due to strong gas drag (Appendix B), the surface number density distribution becomes nearly flat at $r/R_{\rm H}\lesssim0.003$. 

The distribution of planetesimals captured into retrograde orbits can be divided into two regions: the outer region at $r/R_{\rm H} \gtrsim 0.01$, where the distribution can be approximated by a single power-law with $\alpha \simeq 1.2$, and the inner region with nearly flat distribution. 
In the outer region, bodies are captured into relatively long-lived orbits but migrate inward more rapidly than the prograde ones in the same region due to strong headwind. Owing to rapid orbital decay, the distribution levels off inside the critical radius for capture into retrograde orbits by a single encounter (\citealt{F13}; green vertical dashed line).

We also plot the distribution of free planetesimals for comparison. 
Although its surface number density is slightly enhanced due to the effect of the planet's gravitational focusing, 
it is comparable to the surface number density of the background unperturbed planetesimal disk ($n_{\rm s}$).
The surface number density of free planetesimals vanishes inside the prograde capture radius ($r\lesssim0.005R_{\rm H}$), 
because planetesimals become permanently captured in that region due to gas drag.

\subsection{Dependence of Distribution on Planetesimal Size}

\begin{figure}
\epsscale{1.2}
   \plotone{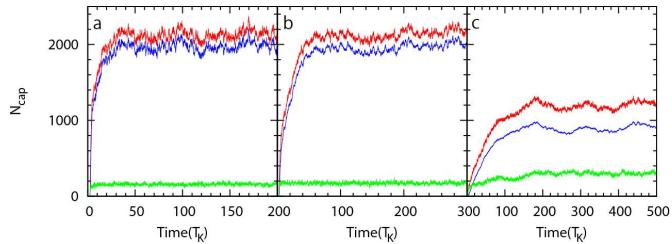}
   \epsscale{1}
 \caption{
Same as Figure~\ref{fig:tncap_6}, but for the cases of (a) $\zeta=10^{-5}$ ($r_{\rm s}\simeq30$ m), 
(b) $\zeta=10^{-7}$ ($r_{\rm s}\simeq3$ km) and (c) $\zeta=10^{-8}$ ($r_{\rm s}\simeq30$ km), respectively.
}
 \label{fig:tncap_578}
\end{figure}

\begin{figure}
\epsscale{1.2}
 \plotone{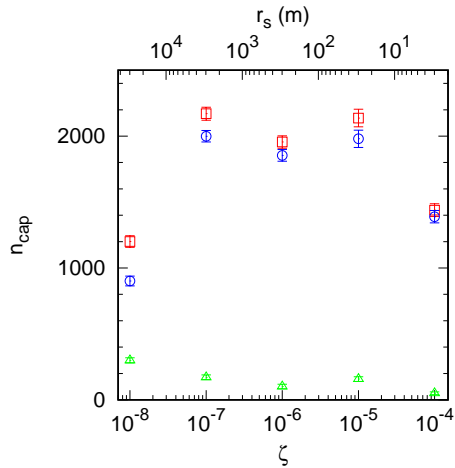}
 \epsscale{1}
 \caption{
Averaged number of captured planetesimals at an equilibrium state as a function of $\zeta$ (or $r_{\rm s}$).
Squares, circles and triangles show the total number of captured planetesimals, and those of prograde and retrograde orbits, respectively.
}
 \label{fig:nave}
\end{figure}

\begin{figure}
\epsscale{1.2}
   \plotone{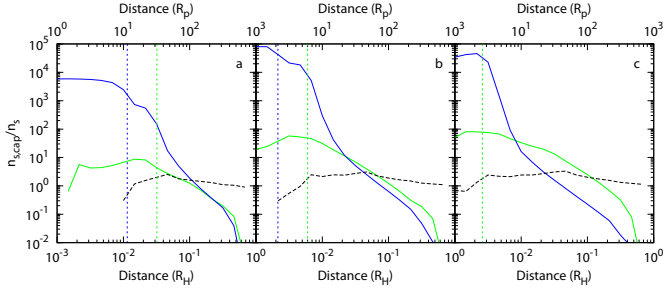}
   \epsscale{1}
 \caption{
 Same as Figure~\ref{fig:rns_6}, but for the cases of (a) $\zeta=10^{-5}$ ($r_{\rm s}\simeq30$ m), (b) $\zeta=10^{-7}$ ($r_{\rm s}\simeq3$ km), 
and (c) $\zeta=10^{-8}$ ($r_{\rm s}\simeq30$ km), respectively.
}
 \label{fig:rns_578}
\end{figure}

Next, we examine the dependence of the distribution of captured planetesimals on the size of planetesimals, 
by changing the values of the gas drag parameter ($\zeta=10^{-8}-10^{-4}$).
Figures~\ref{fig:tncap_578} show the evolution of the number of captured planetesimals for $\zeta=10^{-5}$ (corresponding to $r_{\rm s}\simeq30$ m for the gas-starved disk model), $10^{-7}$ ($r_{\rm s}\simeq3$ km), and $10^{-8}$ $(r_{\rm s}\simeq30$ km), respectively.
General behavior is similar to the case with $\zeta=10^{-6}$ (Figure~\ref{fig:tncap_6}),
but, the timescale to reach the equilibrium state increases with deceasing strength of gas drag because the rates of capture and  orbital decay are reduced.
We find that  the number of planetesimals captured in prograde orbits in the case of $\zeta=10^{-8}$ ($r_{\rm s}\simeq30$ km) is significantly smaller than the other cases, because the capture radius for prograde orbits in this case is smaller than the physical size of the planet.  
On the other hand, the number of planetesimals captured in retrograde orbits in this case is somewhat larger, because the lifetime of planetesimals in retrograde orbits becomes longer under such weak gas drag.
Figure \ref{fig:nave} shows the number of captured planetesimals in the disk in the equilibrium state as a function of $\zeta$ (or $r_{\rm s}$). 
We find that the number hardly depends on the planetesimal size for the intermediate values of $\zeta$, while it decreases for small and large values of $\zeta$ for the reasons described above.
Figure~\ref{fig:rns_578} shows the surface number density distribution for various values of  $\zeta$. 
The general features of the distribution are similar to the case of $\zeta=10^{-6}$ shown in Figure~\ref{fig:rns_6}, which suggests that these features are common for a wide range of gas drag parameters.
In the case of $\zeta=10^{-5}$,  planetesimals can be captured by gas drag even in the outer parts of the disk where the gas
density is rather low. 
Thus, the surface number density of captured planetesimals shifts radially outward. 
On the other hand, when $\zeta=10^{-7}$, planetesimals need to pass through the denser part of the disk in order to be captured,
thus, the distribution shifts radially inward.
In the case of $\zeta=10^{-8}$ shown in Figure~\ref{fig:rns_578}(c),
the inner region corresponding to single-encounter capture into prograde orbits disappears because the prograde capture radius is smaller than the planet's physical radius.

\begin{figure}
\epsscale{1.2}
  \plotone{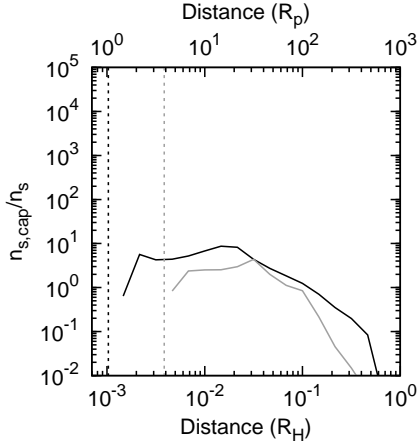}
  \epsscale{1}
 \caption{
Surface number density of planetesimals captured on retrograde orbits as a function of distance from the planet.  
Black and gray lines represent results for $\zeta=10^{-5}$ ($r_{\rm s}\simeq30$ m) and $10^{-4}$ ($r_{\rm s}\simeq3$ m), respectively.
The vertical dotted lines show the analytically-obtained radial distance for the change of the orbital direction \citep{S16} for the above two cases.
}
 \label{fig:rst}
\end{figure}

In some cases,  captured planetesimals change their direction of orbital motion from retrograde to prograde due to strong headwind after they become permanently captured \citep[e.g.,][]{T14, S16}. 
In the case of a planet at Jupiter's orbit, the change of the orbital direction takes place before planetesimals spiral into the planet when $\zeta\gtrsim10^{-5}$ \citep{S16}. 
The above range of $\zeta$ corresponds to  $r_{\rm s}\lesssim10$ m  if we assume the gas density based on the gas-starved disk model \citep{CW02}.
Figure~\ref{fig:rst} shows the surface number density of planetesimals captured into retrograde orbit  as a function of radial distance for the cases with $\zeta=10^{-4}$ and  $10^{-5}$.
The number density vanishes in the vicinity of the planet because planetesimals captured in retrograde orbits change the direction of orbital motion.  
We confirmed that the surface number density of captured planetesimals on prograde orbits increases due to this effect in  the innermost region of the disk when $\zeta=10^{-4}$.

\section{DEPENDENCE OF THE DISTRIBUTION OF CAPTURED PLANETESIMALS ON DYNAMICAL STATES OF INCOMING PLANETESIMALS}
\label{sec:eigap}

So far, we have examined capture of planetesimals that are initially on heliocentric circular orbits with uniform radial distribution in the protoplanetary disk. 
Here, we examine effects of dynamical states of incoming planetesimals, 
i.e., eccentricities and inclinations of their initial heliocentric orbits, and non-uniform radial distribution in the protoplanetary disk.
We assume that eccentricities and inclinations follow a Rayleigh distribution with given rms values (we assume $\langle e_{\rm H}^{2}\rangle^{1/2}=2\langle i_{\rm H}^{2}\rangle^{1/2}$, where $e_{\rm H}$ and $i_{\rm H}$ are the eccentricity and inclination scaled by $R_{\rm H}/a_{0}$), and that orbital phase angles
are randomly distributed in [0-2$\pi$]. 
As for the radial distribution of planetesimals in the protoplanetary disk, a sufficiently massive planet opens a gap both in the gas and solid component of the disk, 
and interactions between solid bodies and the gas near the edge of the gap is important for the supply of solids into the circumplanetary disk  \citep{PM06, R06, A12, DP15}. 
Exact treatment of such interactions requires global simulation of such a system, which is beyond the scope of the present work. 
Instead, in the present work, we adopt a simple model for the non-uniform radial distribution of planetesimals in the protoplanetary disk \citep{F13}, 
by assuming that planetesimals in the vicinity of the planet's orbit have been removed and a gap centered on the planet's orbit with a half width $W_{\rm gap}$ (scaled by $R_{\rm H}$) is formed.

\begin{figure*}
  \plotone{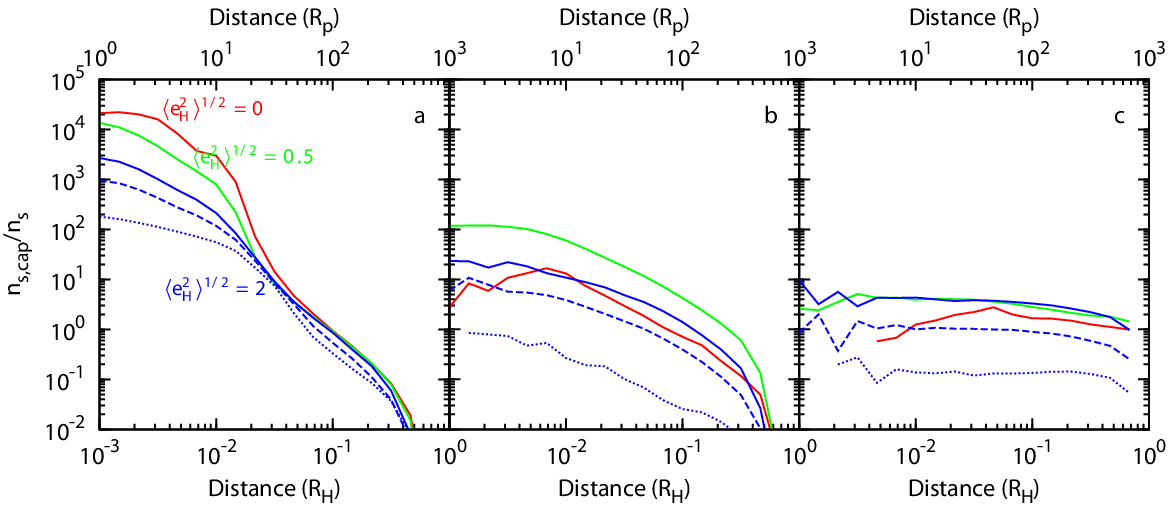}
  \plotone{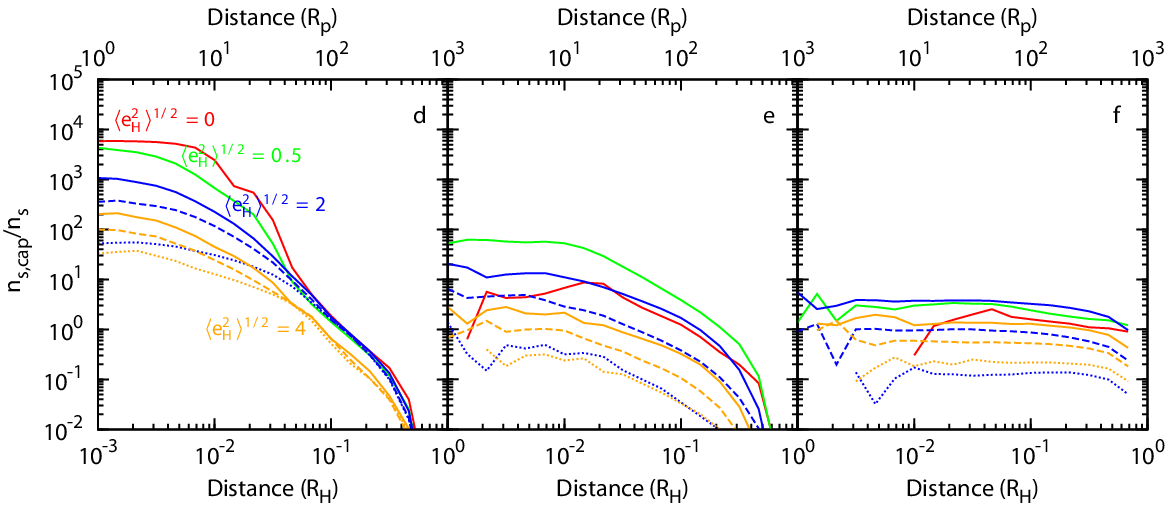}
 \caption{Distribution of the surface number density for the cases of various dynamical states of planetesimal. 
Panels (a) to (c) show the case of $\zeta=10^{-6}$ ($r_{\rm s}\simeq300$ m); (a) the distributions of planetesimals captured in prograde orbits, 
(b) retrograde orbits, and (c) free planetesimals. 
Red, green, and  blue liens show $\langle e_{\rm H}^{2} \rangle^{1/2}=0$, 0.5, and 2, respectively.
The solid line represents the case of uniform radial distribution, 
while the dashed and dotted lines represent the cases with $W_{\rm gap}=3$ and 4, respectively.
Panels (d), (e), and (f) are the same as Panels (a), (b), and (c) but for the cases of $\zeta=10^{-5}$ ($r_{\rm s}\simeq30$ m).
Yellow lines represent results  for $\langle e_{\rm H}^{2} \rangle^{1/2}=4$.}
 \label{fig:com_all6}
\end{figure*}

In Figure~\ref{fig:com_all6}, the solid lines represent results for the cases with uniform radial distribution of planetesimals with different velocity dispersions; $\zeta=10^{-6}$ ($r_{\rm s} \simeq 300$ m) in Panels (a), (b) and (c), while $\zeta = 10^{-5}$ ($r_{\rm s} \simeq 30$ m) in (d), (e), and (f). 
Figures~\ref{fig:com_all6}(a) and (d) show the distribution of planetesimals captured in prograde orbits. 
We find that the surface number density in the inner region that corresponds to capture by a single encounter ($r\lesssim0.02R_{\rm H}$) decreases significantly with increasing velocity dispersion \citep{F13}. 
The bump that we find in the case of initially circular orbits and corresponds to the capture radius is smoothed out in the case with non-zero velocity dispersions, 
because the capture radius depends on the planetesimals' velocity dispersion and those with various eccentricities and inclinations are captured in the cases with non-zero velocity dispersions. 
We find that the dependence of the distribution in the outer region ($r \gtrsim 0.02R_{\rm H}$) on the velocity dispersion is rather weak. 
As we mentioned above, the distribution in this region is determined by a balance between capture into relatively long-lived orbits whose types depend on the velocity dispersion \citep{S11, SO13,S16}, and rather rapid orbital decay due to large eccentricities of planet-centered orbits. 
The resultant enhancement of the surface number density in this region compared to the background value is not large ($n_{\rm s,cap}/n_{\rm s} \lesssim 10$).

Figure~\ref{fig:com_all6}(b) and ~\ref{fig:com_all6}(e) show similar plots for planetesimals captured into retrograde orbits. 
In this case, the surface number density for the case with small but non-zero velocity dispersion is larger than that for the initially circular orbits, because of contribution from long-lived capture orbits that only appear in cases of non-zero velocity dispersions \citep{S16, SO16}. 
With further increase of velocity dispersion, the surface number density decreases for the same reasons in the case of prograde orbits. Figure~\ref{fig:com_all6} (c) and (f) show that free planetesimals exist even in the vicinity of the planet in the case with non-zero velocity dispersions, because limited energy dissipation allows them to avoid permanent capture.

In Figure~\ref{fig:com_all6}, we also plotted with dashed and dotted lines results for the cases with a gap in the planetesimal disk.
In this case, only planetesimals initially on orbits with $|b_{\rm H}| \geq W_{\rm gap}$ can approach the planet and the circumplanetary disk, where $b_{\rm H}=(a-a_{0})/R_{\rm H}$ is  the  difference in the semi-major axes of the planetesimal and the planet scaled by the planet's Hill radius (Figure~\ref{fig:simu}).
Figure~\ref{fig:com_all6}(a) shows that the surface number density of captured bodies on prograde orbits in the inner region decreases due to the effect of the gap, while that in the outer region is hardly affected. Figure~\ref{fig:com_all6}(b) and ~\ref{fig:com_all6}(e) show that the surface number densities of bodies on retrograde orbits as well as free planetesimals decrease significantly; thus the dominance of the bodies on prograde orbits become still notable in the presence of the gap.

\begin{figure}
\epsscale{1.15}
\plotone{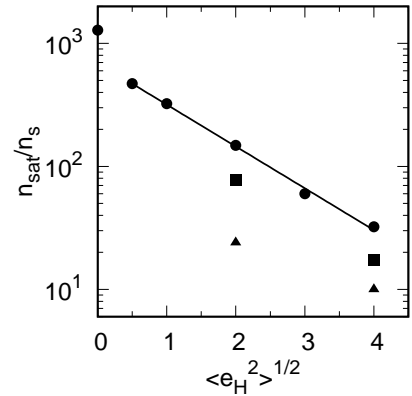}
\epsscale{1}
 \caption{Surface number density of planetesimals captured into prograde orbits averaged over the region of the current radial location of regular satellites ($n_{\rm sat}$) as a function of the velocity dispersion of planetesimals in the case of $\zeta=10^{-5}$ ($r_{\rm s}\simeq30$ m).
Circles, squares, and triangles represent $W_{\rm gap}=0$, 3, 4, respectively.
Solid line is the least-square fit to the results for the case of uniform distribution.}
 \label{fig:e_ns}
\end{figure}

Most of regular satellites of giant planets have orbits with semi-major axes of $\sim0.005R_{\rm H}-0.03R_{\rm H}$. 
Interestingly, our numerical results show that the surface number density of captured planetesimals is significantly enhanced in such regions compared to the background value.
We will further discuss implications of our results for satellite accretion in Section~\ref{sec:dis}.
Using numerical results presented above, here we calculate averaged surface number density of captured planetesimals $n_{\rm sat}$ for the radial region between $r=0.003R_{\rm H}$ and $r=0.03R_{\rm H}$. 
Since planetesimal on prograde orbits are dominant in such a region, we calculate $n_{\rm sat}$ using only planetesimals captured into prograde orbits. 
Figure~\ref{fig:e_ns} shows that $n_{\rm sat}$ decreases with increasing random velocity of planetesimals, regardless of the presence of the gap in the
protoplanetary disk.
Extrapolating on numerical results, 
we find that enhancement of the surface number density of planetesimals captured by gas drag would be negligible in the radial location corresponding to the current orbits of the regular satellites when $\langle e_{\rm H}^{2} \rangle^{1/2}\gtrsim8$.
On the other hand, in the case of non-uniform radial distribution of planetesimals ($W_{\rm gap}>3$),
capture of planetesimals hardly takes place in the shear-dominated regime $\langle e^{2}_{\rm H} \rangle^{1/2}\lesssim1$ because planetesimals approaching the planet are removed by the gap \citep{F13}.  
Therefore, moderate values of the velocity dispersion seem to be preferable for the supply of planetesimals as building blocks of regular satellites in the presence of a gap.


\section{IMPLICATIONS FOR SATELLITE ACCRETION}
\label{sec:dis}
\subsection{Planetesimal-to-Dust Ratio in Circumplanetary Disks}
\label{subsec:ratio}

In the previous sections, we have shown that planetesimals captured from heliocentric orbits are supplied  into the vicinity of  the current orbits of regular satellites. 
Here, we compare the surface density of captured planetesimals with that of dust, in order to examine their relative contribution  to satellite accretion.

\begin{figure}
\epsscale{1.1}
\plottwo{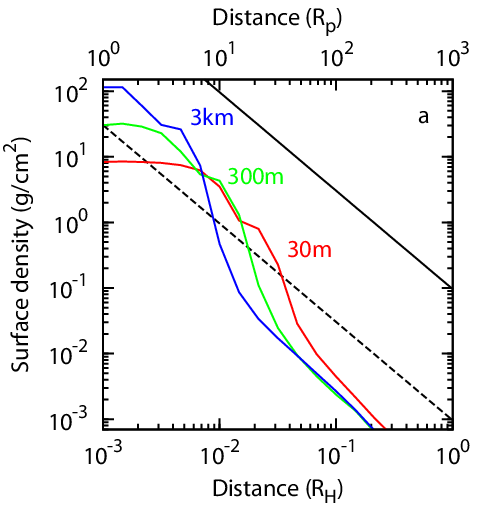}{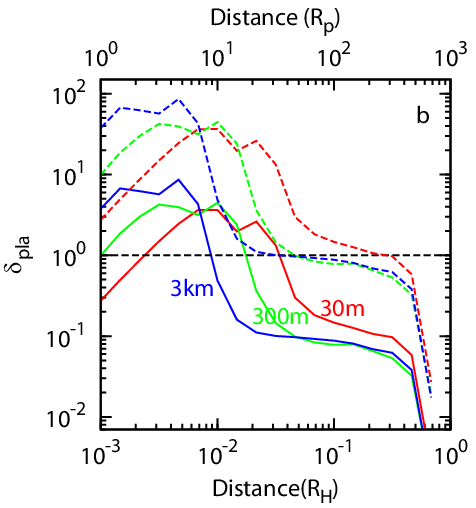}
\epsscale{1}
 \caption{
Panel (a) shows surface density of planetesimals captured from heliocentric circular orbits for several values of $\zeta$ (or $r_{\rm s}$) as a function of the radial distance from the planet.
Red, green, and blue lines represent $\zeta=10^{-5}$ ($r_{\rm s}\simeq30$ m), $\zeta=10^{-6}$ ($r_{\rm s}\simeq300$ m), and $\zeta=10^{-7}$ ($r_{\rm s}\simeq3$ km), respectively.
Solid and dashed lines represent the surface density of gas $\Sigma$ ($\propto r^{-3/2}$) and dust $\Sigma_{\rm dust}$ ($\propto r^{-3/2}$) in the circumplanetary disk, respectively. 
Panel (b) shows the ratio of the surface density of captured planetesimals to that of dust as a function of radial distance from the planet.
Solid and dashed lines show the case with $\chi_{\rm ppd}/\chi=1$ and 10, respectively.
}
 \label{fig:r_sigma_cir}
\end{figure}

We assume that the surface density of the gas in the protoplanetary disk at the time of satellite accretion considered here is given by $\Sigma_{\rm ppd,gas} = f_{\rm dep} \times \Sigma_{\rm MMSN}$, where $\Sigma_{\rm MMSN}$  is the surface density of the minimum mass solar nebula \citep{H81} and $f_{\rm dep}$ is the depletion factor due to disk dispersal. 
We consider a simple model for a circumplanetary disk at the last stage of giant planet formation as follows. 
On the basis of hydrodynamic simulations of gas accretion onto a forming giant planet and the gas-starved model for the circumplanetary disk, 
we set $f_{\rm dep} \simeq 10^{-3}$ (Appendix C). 
As for the circumplanetary disk, we assume here that the power-law distribution of the gas surface density ($\Sigma$) given by Equation~(\ref{eq:snum_svelo}) can be extended to $r=R_{\rm H}$, where it equals $\Sigma_{\rm ppd,gas}$, i.e., $\Sigma = \Sigma_{\rm ppd,gas} (r/R_{\rm H})^{-p}$. 
In this case, if the dust-to-gas ratio in the circumplanetary disk is given by $\chi$, the dust surface density in the circumplanetary disk is given by
\begin{eqnarray}
\Sigma_{\rm dust}&=&\chi\Sigma .
\label{eq:cpd}
\end{eqnarray}
On the other hand, we assume that the ratio of the surface density of planetesimals in the protoplanetary disk ($\Sigma_{\rm ppd,pla}$) to that of the gas is given by $\chi_{\rm ppd}$, i.e., $\Sigma_{\rm ppd,pla} = \chi_{\rm ppd} \Sigma_{\rm ppd,gas}$. 
As for the surface density distribution of planetesimals in the circumplanetary disk, $\Sigma_{\rm pla}(r)$, 
we use our numerical results obtained in the previous sections, assuming that it equals $\Sigma_{\rm ppd,pla}$ at $r=R_{\rm H}$, i.e., $\Sigma_{\rm pla}(R_{\rm H})$ = $\chi_{\rm ppd} \Sigma_{\rm ppd,gas}$. 
Note that we use the total surface number of planetesimals captured into both prograde and retrograde orbits. 
From $\Sigma_{\rm dust}$ and $\Sigma_{\rm pla}$ obtained above, we can calculate the ratio of $\Sigma_{\rm pla}$ to $\Sigma_{\rm dust}$ in the circumplanetary disk as
\begin{eqnarray}
\delta_{\rm pla}=\Sigma_{\rm pla}/\Sigma_{\rm dust},
\end{eqnarray}
as a function of the radial distance $r$ for a given value of $\chi_{\rm ppd}/\chi$.

\begin{figure*}
\epsscale{1.2}
   \plotone{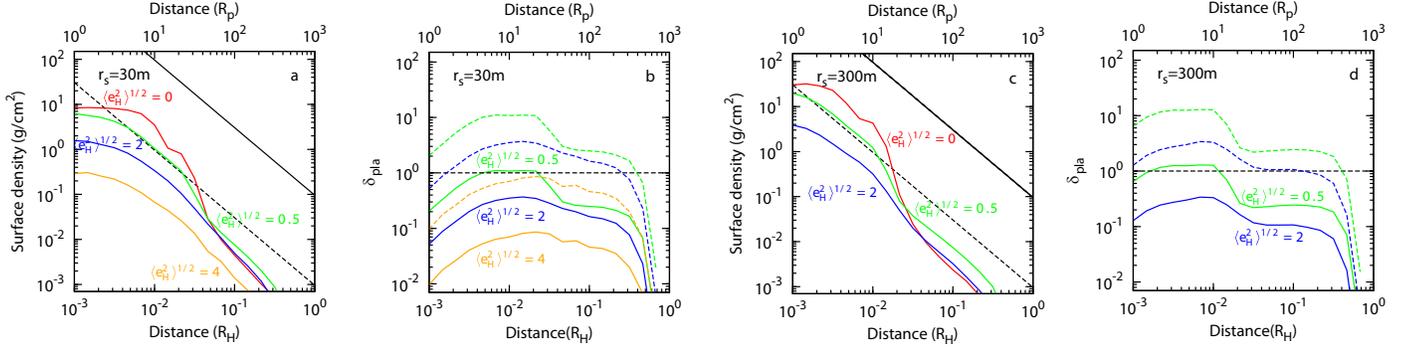}
   \epsscale{1}
 \caption{
Panel (a) shows  surface density for several values of  $\langle e_{\rm H}^{2}\rangle^{1/2}$ in the case of $\zeta=10^{-5}$ ($r_{\rm s}\simeq30$ m).
Red, green, blue, and yellow lines represent $\langle e_{\rm H}^{2}\rangle^{1/2}=0$, 0.5, 2, and 4, respectively.
Panel (b) shows  planetesimal-dust ratio for the cases shown in Panel (a).
Panels (c) and (d) are the same as Panels (a) and (b), but for the case of  $\zeta=10^{-6}$ ($r_{\rm s}\simeq300$ m).
}
\label{fig:r_sigma_ecc}
\end{figure*}

Figure~\ref{fig:r_sigma_cir} shows the surface density of the gas (solid straight line), dust (dashed straight line), and captured planetesimals (curves) for the case of  initially on heliocentric circular orbits with $\chi_{\rm ppd}=\chi=0.01$.
We find that solid material in the disk is dominated by captured planetesimals in the inner region, while dust particles are dominant in the outer region.
The range of the region dominated by captured planetesimals shift inward  with decreasing gas drag parameter (i.e., increasing size of planetesimals).
Figure~\ref{fig:r_sigma_cir}(b) shows the planetesimal-to-dust ratio as a function of the radial distance, where we can confirm the above tendency. 
In this figure, we also plot the case with $\chi_{\rm ppd}/\chi=10$ ($\chi_{\rm ppd}=5\times10^{-2}$ and $\chi=5\times10^{-3}$), 
because the dust abundance in the circumplanetary disk would be reduced by, for example, the filtering effect due to accumulation at the gas density bump in the protoplanetary disk \citep{A12}. 
In this case, the planetesimal-dominated region becomes extended almost to the outer edge of the circumplanetary disk ($r\lesssim0.2R_{\rm H}$).
However, the planetesimal-dominated region shrinks when incoming planetesimals have non-zero velocity dispersions in the protoplanetary disk (Figure~\ref{fig:r_sigma_ecc}). 
The region further shrinks if there is a wide gap in the radial distribution of planetesimals in the protoplanetary disk (Figure~\ref{fig:com_all6}).
From these results, we conclude that the contribution of captured planetesimals would become comparable to or even larger than that of dust particles as building blocks of regular satellites when the velocity dispersion  of planetesimals and the width of their gap in the protoplanetary disk are rather small, while their contribution becomes small in the case of larger velocity dispersions and/or a wide gap.

\subsection{Timescale of Accretion of Captured Planetesimals}
Using the surface density distribution of captured planetesimals that we obtained in the previous sections,  
we can evaluate the timescale of collision among them and examine whether they grow by accretion before spiraling into the planet.
For this purpose, first, we estimate the relative velocity between captured planetesimals, assuming that it is determined by damping due to gas drag and enhancement due to viscus stirring among the planetesimals.
The timescale of velocity damping by gas drag is given by
\begin{eqnarray}
T_{\rm damp}=m_{\rm s}u/F_{\rm drag}=\frac{2m_{\rm s}h}{C_{\rm D}\pi r^{2}_{\rm s}\Sigma u},
\end{eqnarray}
where $u\simeq(e_{\rm p}^{2}+i_{\rm p}^{2}+\eta^{2})^{1/2}v_{\rm K}$ is the relative velocity between the gas and planetesimals 
($i_{\rm p}$ is  an inclination of the planet-centered orbit), and we used $\rho_{\rm gas}\simeq \Sigma/h$. 
On the other hand, the timescale of viscus stirring among captured planetesimals is given by
\begin{eqnarray}
T_{\rm vs}=\frac{1}{ n_{\rm pla}\pi r_{\rm g}^{2} v},
\end{eqnarray}
where $n_{\rm pla}$ is number density of captured planetesimals in the circumplanetary disk, 
$r_{\rm g}=2Gm_{\rm s}/v^{2}$ is the impact parameter for the 90$^\circ$ deflection, and $v=(e_{\rm p}^{2}+i_{\rm p}^{2})^{1/2}v_{\rm K}$. 
Since $n_{\rm pla}$ can be approximately given by $n_{\rm pla}\simeq n_{\rm s, cap}\Omega_{\rm p}/v$,  $T_{\rm vs}$ is rewritten as
\begin{eqnarray}
T_{\rm vs}=\frac{v^{4}}{4G^{2}m_{\rm s}^{2}\pi n_{\rm s,cap} \Omega_{\rm p}{\rm ln}\Lambda},
\end{eqnarray}
where ${\rm ln}\Lambda$ comes from the effect of distant encounters, and ${\rm ln}\Lambda\sim1$ in the dispersion-dominated regime \citep{O02}.
Setting $T_{\rm damp}=T_{\rm vs}$, the equilibrium velocity dispersion $v_{\rm eq}$ is obtained as \citep{KI00, K10}
\begin{eqnarray}
\label{eq:veq}
v_{\rm eq}&=&\left(\frac{2{\rm ln}\Lambda}{C_{\rm D}}\right)^{1/5}\left(\frac{\Sigma_{\rm pla}}{\Sigma}\right)^{1/5}\left( h\Omega_{\rm p}\right)^{1/5}v_{\rm esc}^{4/5} ,
\end{eqnarray}
where $v_{\rm esc}=\sqrt{2Gm_{\rm s}/r_{\rm s}}$ is the escape velocity of planetesimals, and we have neglected the relative velocity due to the pressure gradient. 

Using this equilibrium velocity dispersion, the timescale of collision between planetesimals is written as
\begin{eqnarray}
T_{\rm col}&=&\frac{1}{n_{\rm pla}\sigma_{\rm col} v_{\rm eq}}=\frac{1}{n_{\rm s,cap}\sigma_{\rm col}\Omega_{\rm p}}.
 \end{eqnarray}
Substituting $\sigma_{\rm col}=\pi r_{\rm s}^{2}\left(1+v_{\rm esc}^{2}/v_{\rm eq}^{2}\right)$ for the collision cross section, we have
\begin{eqnarray}
T_{\rm col}/T_{K}=\frac{1}{2n_{\rm s,cap} \pi^{2} r_{\rm s}^{2}\left(1+v_{\rm esc}^{2}/v_{\rm eq}^{2}\right)}\left(\frac{\Omega_{\rm p}}{\Omega}\right)^{-1}.
\end{eqnarray}
On the other hand, the timescale of the orbital decay due to gas drag can be written as \citep{A76}
\begin{eqnarray}
T_{\rm decay}/T_{\rm K}=\frac{1}{2\pi\eta}\frac{1+g^{2}}{2g}\left(\frac{\Omega_{\rm p}}{\Omega}\right)^{-1},
\end{eqnarray}
where
\begin{eqnarray}
g=a_{\rm drag}/(u\Omega_{\rm p})=\zeta \eta \tilde{r}^{-\gamma+1}.
\end{eqnarray}

\begin{figure*}
\epsscale{1.18}
  \plotone{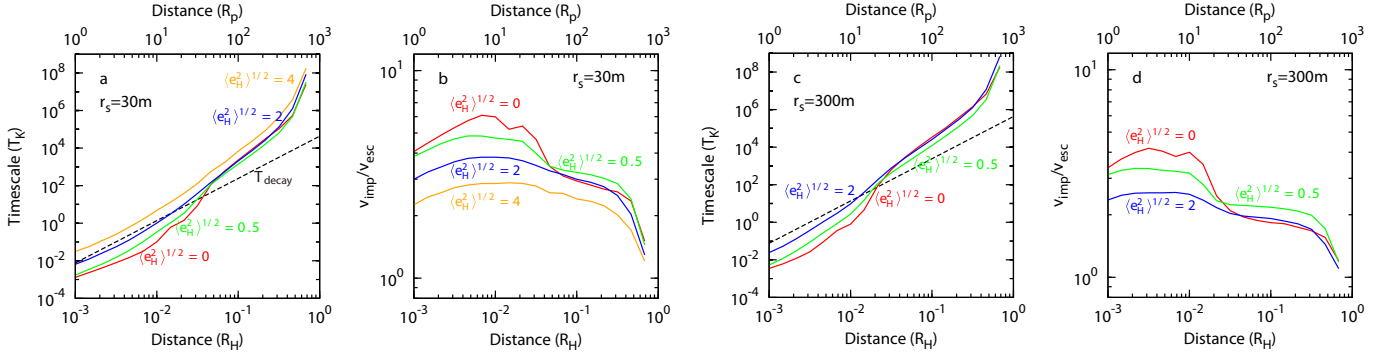}
  \epsscale{1}
 \caption{Panel (a) shows collision timescale for several values of $\langle e_{\rm H}^{2}\rangle^{1/2}$  in the case of $\zeta=10^{-5}$ ($r_{\rm s}\simeq30$ m).
Red, green, blue, and yellow lines represent $\langle e_{\rm H}^{2}\rangle^{1/2}=0$, 0.5, 2, and 4, respectively.
Dashed line represents timescale of orbital decay due to gas drag.
Panel (b) shows impact velocity between captured planetesimals in the circumplanetary disk  for several values of $\langle e_{\rm H}^{2}\rangle^{1/2}$ as a function of the radial distance from the planet.
Panels (c) and (d) are the same as Panels (a) and (b), but for the case of  $\zeta=10^{-6}$ ($r_{\rm s}\simeq300$ m).
}
 \label{fig:tcol}
\end{figure*}

Figure~\ref{fig:tcol}(a) shows $T_{\rm col}$ and $T_{\rm decay}$ as a function of the radial distance from the planet in the case of $\zeta=10^{-5}$ ($r_{\rm s}\simeq30$m).
In the outer region of the circumplanetary disk with $r\gtrsim0.04R_{\rm H}$, $T_{\rm decay}<T_{\rm col}$,
and  captured planetesimals migrate inward  before they grow by accretion.
On the other hand, in the inner region where the surface density rapidly increases ($r\lesssim0.03R_{\rm H}$; Figure~\ref{fig:com_all6}), $T_{\rm col} < T_{\rm decay}$ 
and their mutual collision becomes important.

Figure~\ref{fig:tcol}(b) shows the impact velocity $(v_{\rm imp}=\left(v_{\rm eq}^{2}+v_{\rm esc}^{2}\right)^{1/2})$ between captured planetesimals as a function of the radial distance in the above case.
Although the impact velocity is higher than the escape velocity, it is rather low ($\sim 0.01$ ms$^{-1}$).
Recent impact simulations with the effect of material strength show that bodies can accrete and grow in such low-velocity collisions \citep{JA15}.
Thus, we can expect collisional growth of captured planetesimals in the inner region of the circumplanetary disk. 
On the other hand,  the surface density of captured planetesimals largely depends on the random velocity of the planetesimals disk.
As a result, $T_{\rm col}$ increases with increasing velocity dispersion, 
and is comparable to or larger than $T_{\rm decay}$ when $\langle e_{\rm H}^{2} \rangle^{1/2}=2-4$, even in the inner part of the disk. 
In this case, captured planetesimals would spiral into the planet before they  grow significantly by accretion.

In the case of  $\zeta=10^{-6}$ ($r_{\rm s}\simeq300$ m; Figure~\ref{fig:tcol}(c)),
the collision timescale in the inner region of the disk becomes somewhat longer than the case with $\zeta=10^{-5}$ ($r_{\rm s}\simeq30$ m).
The velocity dispersion becomes lower also due to the lower surface density of captured planetesimals (Figure~\ref{fig:tcol}(d)).
On the other hand, the orbital decay timescale becomes longer, because of weaker gas drag. 
As a result, $T_{\rm col}$  becomes shorter than $T_{\rm decay}$ in the vicinity of the planet in the inner region of the disk, and growth of planetesimals 
by accretion can be expected there.
As we mentioned in Section~\ref{sec:eigap}, the surface number density of captured planetesimals is significantly enhanced in the regions corresponding to the current radial location of regular satellites of giant planets.
Our results  suggest that once a significant amount of planetesimals are captured in such regions, they would contribute to the growth of regular satellites of giant planets.
Further studies including mutual gravitational interactions between captured planetesimals are desirable to quantitatively examine their effect on the growth of satellites.


\section{SUMMARY}
\label{sec:sum}

In the present work, we examined the distribution of planetesimals captured in circumplanetary disks using orbital integration. 
We found that the number of captured planetesimals in the circumplanetary disk reaches an equilibrium state as a balance between continuous capture from their heliocentric orbits by gas drag and orbital decay into the planet. 
Typically, the number of planetesimals captured into retrograde orbits is much smaller than that on prograde orbits, 
because those captured on retrograde orbits experience strong headwind and spiral into the planet rapidly.
Using our numerical results, we obtained surface number density distribution of permanently captured planetesimals in the circumplanetary disk in the equilibrium state. 
We found that the surface number density of captured planetesimals is significantly enhanced at the current location of regular satellites, and that it depends on the dynamical states of planetesimals, 
i.e., their initial velocity dispersion in the protoplanetary disk and the presence of a gap in their radial distribution; larger velocity dispersions and a wider gap both reduce the efficiency of planetesimal capture, thus decreases their surface number density in the disk.

We also examined the surface density ratio of dust and captured planetesimals in the circumplanetary disk.
We found that solid material at the current location of regular satellites is dominated by captured planetesimals when the random velocity of planetesimals is rather small and a wide gap is not formed in the protoplanetary disk. 
In this case, captured planetesimals can grow by mutual collision before they spiral into the planet due to gas drag.
Therefore, captured planetesimals would significantly contribute to the growth of regular satellites.
Further studies including mutual gravitational interactions between captured planetesimals are necessary to quantitatively examine their effect on satellite accretion. 
Also, their importance on satellite accretion relative to dust particles depends on dust-to-planetesimal ratio in the protoplanetary disk and in the circumplanetary disk.
Although, the influence of the material density on the radial distribution of captured bodies is expected to be small,
mass of captured planetesimals would be reduced by ablation during the passage through the circumplanetary disk \citep{F13, DP15, SO16}.
For example, in \citet{SO16}, we examined the effect of ablation during capture of planetesimals by a waning circumplanetary gas disk. 
We found that the effect of ablation in the case of planetesimals captured into planet-centered orbits with large semi-major axes ($\gtrsim 0.1 R_{\rm H}$) is insignificant, 
because the mass loss due to ablation mostly occurs during a short period of time when the bodies pass through the dense part of the circumplanetary  disk near their pericenter. 
On the other hand, when their semi-major axes become sufficiently small due to orbital decay by gas drag, 
the captured planetesimals constantly pass through the dense part of the disk, and their mass loss would be significant.
Further studies of the influence of ablation of captured planetesimals on satellite accretion would also be desirable.


\acknowledgments
We thank Takayuki Tanigawa and Toshihiko Kadono for discussion, and the anonymous reviewer for useful comments on the manuscript. 
This work was supported by JSPS Grants-in-Aid for JSPS Fellows (12J01826) and Scientific Research B (22340125 and 15H03716).
Part of numerical calculations were performed using computer systems at the National Astronomical Observatory of Japan.

\appendix

\section*{APPENDIX A. \\
Gas Drag Coefficient for Small Planetesimals}

In the present work, we examine cases with $r_{\rm s}=1-10^{4}$ m. 
The assumption of $C_{\rm D}=1$ is reasonable for large planetesimals, 
but the dependence of $C_{\rm D}$ on the Mach number and the Reynolds number would be important for small bodies. 
In order to check the validity of the above assumption, we examined values of the drag coefficient of planetesimals orbiting in the circumplanetary gas disk based on the gas-starved disk model \citep{CW02}, and taking account of its dependence on the Mach number and the Reynolds number following \citet{T14}. 
For planetesimals captured into retrograde orbits, we found that $C_{\rm D}\simeq2$ regardless of their sizes, 
because of their large velocity relative to the gas. For those captured into prograde orbits, $C_{\rm D}$ can become as large as $5-7$ in the outer part of the disk with $r/R_{\rm H} \gtrsim 0.1$, but the values become smaller and close to unity in the inner part of the disk; even in the case of $r_{\rm s} = 1$m, we found $1 \lesssim C_{\rm D} \lesssim 3$ at $r/R_{\rm H} \lesssim 0.02$, which corresponds to the current radial locations of the Galilean satellites. 
Therefore, we think that the above assumption is reasonable for the size range considered in the present work, 
although a more exact treatment is required for still smaller bodies.

\section*{APPENDIX B. \\
Orbital  Evolution After Capture}

\begin{figure*}
\epsscale{1.2}
  \plotone{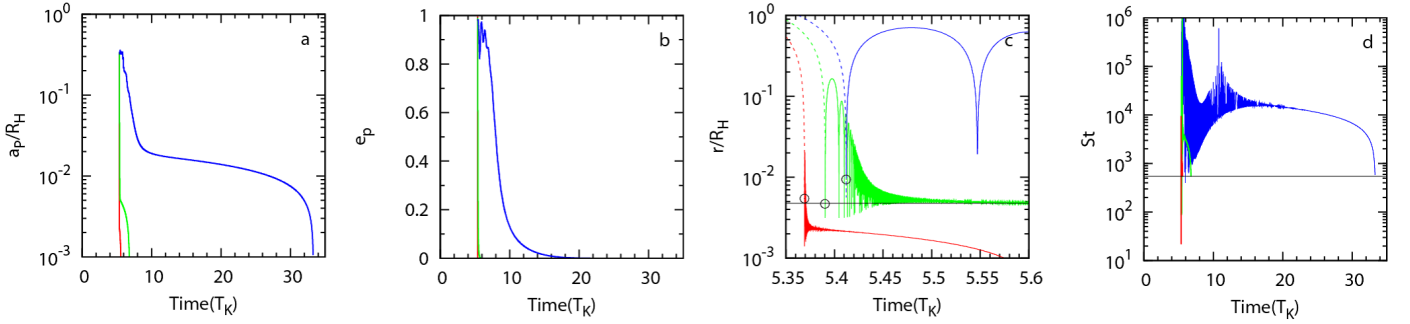}
  \epsscale{1}
  \caption{
Time evolution of semi-major axes (a) and eccentricities (b) of planet-centered orbits, radial distances from the planet (c), and the Stokes number (d) of
captured planetesimals for the cases shown in  Figure~\ref{fig:cap_orbit}.
The dashed lines in panel (c) represent the phase before capture $(E>0)$, 
and the solid lines show the evolution after permanent capture $(E<0)$.
The open circles in Panel (c) show the radial location where the planetesimal becomes permanently captured.
The horizontal line in Panel (c) shows the prograde capture radius \citep[see,][]{F13}. 
In Panel (d), the horizontal line shows the analytically-obtained Stokes number ($\simeq544$) for orbits whose size is the same as the physical radius of the planet \citep{S16}. 
}
\label{fig:time_aer}
\end{figure*}

In the case of capture of planetesimals  initially on heliocentric circular orbits into planet-centered prograde orbits shown in Figure~\ref{fig:tncap_6},  the number of captured planetesimals rapidly increases for $0\lesssim t/T_{\rm K}\lesssim 5$, and  its growth slows down for $5\lesssim t/T_{\rm K}\lesssim 35$. 
This can be explained by the difference in the manner of orbital decay after capture.
Figures~\ref{fig:cap_orbit} show orbital behavior of planetesimals captured from heliocentric circular orbits  when $\zeta=10^{-6}$, for three cases with slightly different initial conditions. 
The red and green lines show the case of rapid orbital decay after permanent capture, while the blue line shows the case of slow decay. 
The time evolution of several quantities for these orbits are shown in Figure~\ref{fig:time_aer}.
In the case of long-lived orbits shown by the blue line (Figures~\ref{fig:cap_orbit}, \ref{fig:time_aer}(a) and \ref{fig:time_aer}(b)), 
first, semi-major axis of the planet-centered orbit $a_{\rm p}$ decreases rapidly, then the body experiences a phase of rather slow orbital decay before spiraling into the planet.
This is because the body maintains a rather large pericenter distance after permanently captured, in contrast to the case of the other short-lived orbits (Figure~\ref{fig:time_aer}(c)).
Figure~\ref{fig:time_aer}(d) shows the evolution of the Stokes number $(St\equiv u/(a_{\rm drag}\Omega_{\rm p}^{-1}))$. 
The  evolution of the Stokes number strongly depends on the eccentricity of the planet-centered orbit, $e_{\rm p}$. 
When $e_{\rm p}$ is large,  the radial region swept by the planetesimal is very wide, and the relative velocity between the planetesimal and the gas significantly changes depending on the distance from the planet. 
This causes the large variation of the Stokes number in Figure~\ref{fig:time_aer}(d).
After $e_{\rm p}$ becomes sufficiently small, the oscillation in the Stokes number ceases because the radial distance from the planet becomes nearly constant. 

\section*{APPENDIX C. \\
A Simple Model for the Circumplanetary Disk at the Last Stage of Giant Planet Formation}

Hydrodynamic simulations of gas accretion onto a forming giant planet show the formation of circumplanetary disk, 
whose gas density distribution is nearly axisymmetric in the vicinity of the planet. 
Although detailed structure of the disk depends on the parameters adopted in simulations, 
here we consider a simple model on the basis of recent high resolution simulations \citep{M08, T12}. 
These simulations show that the size of the circumplanetary disk is given by $r_{\rm d} \simeq 0.2R_{\rm H}$ with $R_{\rm H}$ being the planet's Hill radius, 
and the gas surface density at this disk edge is approximately 10 times enhanced compared to the unperturbed surface density of the protoplanetary disk, $\Sigma_{\rm ppd,gas}$. 
We assume that $\Sigma_{\rm ppd,gas}$ at this last stage of giant planet formation is depleted by a factor $f_{\rm dep}$ compared to the gas surface density of the minimum mass solar nebula model, $\Sigma_{\rm MMSN}$; $\Sigma_{\rm ppd,gas} = f_{\rm dep} \Sigma_{\rm MMSN}$. 
On the other hand, the typical value of the gas surface density at the outer edge of the circumplanetary disk in the gas-starved disks model is about 1 gcm$^{-2}$ \citep{CW02}, 
while the gas surface density at Jupiter's orbit is on the order of 10$^{2}$ gcm$^{-2}$ in the minimum mass solar nebula model. 
Therefore, if $f_{\rm dep} \simeq 10^{-3}$,  the gas surface density of the protoplanetary disk at Jupiter's orbit at the time of satellite accretion is $\Sigma_{\rm ppd,gas} \simeq 10^{-1}$ gcm$^{-2}$, which gives the surface density of the circumplanetary disk roughly consistent with the gas-starved disk model.
\\


%

\begin{thebibliography}{}
\bibitem[Adachi et al.(1976)]{A76}Adachi I., Hayashi, C., \& Nakazawa, K. 1976, Prog. Ther. Phys., 56, 1756
\bibitem[Ayliffe et al.(2012)]{A12} Ayliffe, B. A., Laibe, G., Price, D. J., \& Bate, M. R.  2012, \mnras, 423, 1450
\bibitem[Canup \& Ward(2002)]{CW02}Canup, R. M., \& Ward, W. R. 2002, \aj, 124, 3404
\bibitem[Canup \& Ward(2006)]{CW06}Canup, R. M., \& Ward, W. R. 2006, \nat, 441, 834
\bibitem[Canup \& Ward(2009)]{CW09}Canup, R. M., \& Ward, W. R. 2009, in Europa (Tucson, AZ: Univ. Arizona Press), 59
\bibitem[D'Angelo \& Podolak(2015)]{DP15}D'Angelo, G. \& Podolak, M. 2015, \apj, 806, 203
\bibitem[Dwyer et al.(2013)]{D13}Dwyer, C. A., Nimmo, F., Ogihara, M., \& Ida, S. 2013, \icarus, 225, 390
\bibitem[Estrada et al.(2009)]{E09} Estrada, P. R., Mosqueira, I., Lissauer, J. J., D'Angelo, G., \& Cruikshank, D. P.  2009, Europa (Tucson, AZ:Univ. Arizona Press), 27
\bibitem[Fujita et al.(2013)]{F13}Fujita T., Ohtsuki K., Tanigawa T., Suetsugu R. 2013 \aj, 146, 140
\bibitem[Hayashi(1981)]{H81}Hayashi, C. 1981, Prog. Theor. Phys. Suppl., 70, 35\
\bibitem[Jutzi \& Asphaug(2015)]{JA15}Jutzi, M., \& Asphaug, E. 2015, Science, 348, 1355
\bibitem[Kobayashi et al.(2010)]{K10}Kobayashi, H., Tanaka, H., Krivov, A. V., \& Inaba, S. 2010, \icarus, 209, 836
\bibitem[Kobayashi(2015)]{K15}Kobayashi, H. 2015, Earth Planets Space, 67, 60
\bibitem[Kokubo \& Ida(2000)]{KI00}Kokubo, E., Ida, S. 2000, \icarus, 143, 15
\bibitem[Miguel \& Ida(2016)]{MI16}Miguel, Y., Ida, S. 2016, \icarus, 266, 1 
\bibitem[Mosqueira \& Estrada(2003)]{ME03}Mosqueira, I., \& Estrada, P. R. 2003, Icarus, 163, 198
\bibitem[Machida et al.(2008)]{M08} Machida, M. N., Kokubo, E., Inutsuka, S., \& Matsumoto, T.  2008, \apj, 685, 1220
\bibitem[Ogihara \& Ida(2012)]{OI12}Ogihara, M., \& Ida, S. 2012, \apj, 753, 60
\bibitem[Ohtsuki et al.(2002)]{O02}Ohtsuki, K., Stewart, G. R., \& Ida, S. 2002, \icarus, 155, 436 
\bibitem[Ohtsuki(2012)]{O12}Ohtsuki, K. 2012, Prog. Theor. Phys. Suppl., 195, 29
\bibitem[Paardekooper \& Mellema(2006)]{PM06} Paardekooper, S.-J., \& Mellema, G.  2006, \aap, 453, 1129
\bibitem[Rice et al.(2006)]{R06} Rice, W. K. M., Armitage, P. J., Wood, K., \& Lodato, G.  2006, \mnras, 373, 1619
\bibitem[Sasaki et al.(2010)]{S10}Sasaki, T., Stewart, G, R., Ida, S. 2010 \apj, 714, 1052
\bibitem[Sekine \& Genda(2012)]{SG12}Sekine, Y., \& Genda, H. 2012, Planet. Space Sci. 63, 133
\bibitem[Suetsugu et al.(2011)]{S11}Suetsugu R., Ohtsuki K., Tanigawa T. 2011, \aj, 142, 200 
\bibitem[Suetsugu \& Ohtsuki(2013)]{SO13}Suetsugu, R., Ohtsuki, K. 2013,\mnras, 431, 1709
\bibitem[Suetsugu \& Ohtsuki(2016)]{SO16}Suetsugu, R., Ohtsuki, K. 2016, \apj, 820, 128 
\bibitem[Suetsugu et al.(2016)]{S16}Suetsugu, R., Ohtsuki, K., Fujita, T. 2016 \aj, 151, 140
\bibitem[Tanaka et al.(2002)]{T02} Tanaka, H., Takeuchi, T., \& Ward, W. R. 2002, \apj, 565, 1257
\bibitem[Tanigawa et al.(2014)]{T14} Tanigawa, T., Maruta, A., \& Machida, M. N.  2014, \apj, 784, 109 
\bibitem[Tanigawa et al.(2012)]{T12} Tanigawa, T., Ohtsuki, K., \& Machida, M. N.  2012, \apj, 747, 47

\end{thebibliography}
\end{document}